\documentclass[10pt, conference, compsocconf]{IEEEtran}
\ifCLASSINFOpdf
\else
\fi
\usepackage[tight,footnotesize]{subfigure}

\usepackage[caption=false,font=footnotesize]{subfig}

\usepackage[T1]{fontenc}
\usepackage[utf8]{inputenc}
\usepackage[english]{babel}

\usepackage{calc}
\usepackage{amsmath}
\usepackage{amssymb}
\usepackage{amstext}
\usepackage{mathrsfs}
\usepackage{amsfonts}
\usepackage{dsfont} 
\usepackage{stmaryrd}

\usepackage{color}
\usepackage{multirow}
\usepackage{array}
\usepackage{graphicx}
\usepackage{url}

\usepackage[english]{varioref}
\usepackage[ruled,vlined]{algorithm2e}

\usepackage[amsmath,thmmarks]{ntheorem}

\theoremsymbol{\ensuremath{\Box}}
\theoremstyle{nonumberplain}

\theoremsymbol{\ensuremath{\Box}}
\theoremseparator{:}
\theoremstyle{plain}
\newtheorem{definition}{Definition}
\newtheorem{notation}{Notation}

\newtheorem{example}{Example}
\newtheorem{remark}{Remark}

\usepackage{color}

\begin{document}
%
\title{A Robust Data Hiding Process Contributing to the Development of a Semantic Web}



\author{\IEEEauthorblockN{Jacques M. Bahi, Jean-Fran\c cois Couchot, Nicolas
Friot, and  Christophe
Guyeux}

\IEEEauthorblockA{FEMTO-ST Institute, UMR 6174 CNRS\\
Computer Science Laboratory DISC\\
University of Franche-Comt\'{e}\\
Besan\c con, France\\
\{jacques.bahi, jean-francois.couchot, nicolas.friot,
christophe.guyeux\}@femto-st.fr}\\ 
}


%


\maketitle

\begin{abstract}
%
%
In this paper, a novel steganographic scheme based on
chaotic iterations is proposed. This research work takes place into the information hiding framework,
and focus more specifically on robust steganography.
Steganographic algorithms can participate in the development of a semantic web:
medias being on the Internet can be enriched by information related to their contents, authors, etc., 
leading to better results for the search engines that can deal with such tags. 
As media can be modified by users for various reasons, it is preferable that 
these embedding tags can resist to changes resulting from some classical
transformations as for example cropping, rotation, image conversion, and so on.
This is why a new robust watermarking scheme for semantic search engines is proposed in this document.
For the sake of completeness, the robustness of this scheme is finally 
compared to existing established algorithms.
\end{abstract}

\begin{IEEEkeywords}
\textit{Semantic Web}; \textit{Information Hiding}; \textit{Steganography}; \textit{Robustness}; \textit{Chaotic Iterations}.
\end{IEEEkeywords}

%
\IEEEpeerreviewmaketitle

\section{Introduction}\label{sec:intro}

%

Social search engines are frequently presented as a next generation approach to query the world wide web.
In this conception, contents like pictures or movies are tagged with descriptive
labels by contributors, and search results are enriched with these
descriptions. These collaborative taggings, used for example in Flickr~\cite{Frick} and Delicious~\cite{Delicious} websites, can participate to the development of a Semantic Web, in which every Web page contains machine-readable metadata that describe its content. To achieve this goal by embedding such metadata, information hiding technologies can be useful. 
Indeed, the interest to use such technologies lays on the possibility to realize  social search without websites and databases: descriptions are directly embedded into media, whatever their formats. 

In the context of this article, the problem consists in embedding tags into
internet medias, such that these tags persist even after user transformations.
Robustness of the chosen watermarking scheme is thus required in this situation,
as descriptions should resist to user modifications like resizing,
compression, and format conversion or other classical user transformations in
the field. Indeed, quoting Kalker in~\cite{Kalker2001}, ``Robust watermarking is
a mechanism to create a communication channel that is multiplexed into original content [...]
It is required that, firstly, the perceptual degradation of the marked content
[...] is minimal and, secondly, that the capacity of the watermark channel
degrades as a smooth function of the degradation of the marked content''. The
development of social web search engines can thus be strengthened by the design
of robust information hiding schemes. Having this goal in mind, we explain in
this article how to set up a secret communication channel using a new
 robust steganographic process called $\mathcal{DI}_3$. This new scheme has been
 theoretically presented in~\cite{ihtiap-2012} with an evaluation of its
 security. So, the main objective of this work
 is to focus on robustness aspects presenting firstly  other known schemes in
 the literature, and presenting secondly this new  scheme and and evaluate its
 robustness. This article is thus a first  work on the subject, and the
 comparison with other  schemes concerning the robustness will be realized in
 future work.

The remainder of this document is organized as follows. In Section
\ref{sec:msc-lsc}, some basic reminders concerning the notion of Most and Least
Significant Coefficients are given. In Section~\ref{stateArt}, some well-known steganographic schemes are recalled,
 namely the YASS~\cite{DBLP:conf/ih/SolankiSM07},
 nsF5~\cite{DBLP:conf/mmsec/FridrichPK07}, MMx~\cite{DBLP:conf/ih/KimDR06}, and
 HUGO~\cite{DBLP:conf/ih/PevnyFB10} algorithms. In the next section the
 implementation of the steganographic process $\mathcal{DI}_3$ is detailed,
 and its robustness study is exposed in Section~\ref{sec:robustness-study}.
 This research work ends by a conclusion section, where our contribution is summarized and intended future researches are presented.

\section{Most and Least Significant
Coefficients}\label{sec:msc-lsc}

We first notice that terms of the original content $x$ that may be replaced by terms issued
from the watermark $y$ are less important than others: they could be changed 
without be perceived as such. More generally, a 
\emph{signification function} 
attaches a weight to each term defining a digital media,
depending on its position $t$.

\begin{definition}
A \emph{signification function} is a real sequence 
$(u^k)^{k \in \mathds{N}}$. 
\end{definition}

\begin{example}\label{Exemple LSC}
Let us consider a set of    
grayscale images stored into portable graymap format (P3-PGM):
each pixel ranges between 256 gray levels, \textit{i.e.},
is memorized with eight bits.
In that context, we consider 
$u^k = 8 - (k  \mod  8)$  to be the $k$-th term of a signification function 
$(u^k)^{k \in \mathds{N}}$. 
Intuitively, in each group of eight bits (\textit{i.e.}, for each pixel) 
the first bit has an importance equal to 8, whereas the last bit has an
importance equal to 1. This is compliant with the idea that
changing the first bit affects more the image than changing the last one.
\end{example}

\begin{definition}
\label{def:msc-lsc}
Let $(u^k)^{k \in \mathds{N}}$ be a signification function, 
$m$ and $M$ be two reals s.t. $m < M$. 
\begin{itemize}
\item The \emph{most significant coefficients (MSCs)} of $x$ is the finite 
  vector  $$u_M = \left( k ~ \big|~ k \in \mathds{N} \textrm{ and } u^k 
    \geqslant M \textrm{ and }  k \le \mid x \mid \right);$$
 \item The \emph{least significant coefficients (LSCs)} of $x$ is the 
finite vector 
$$u_m = \left( k ~ \big|~ k \in \mathds{N} \textrm{ and } u^k 
  \le m \textrm{ and }  k \le \mid x \mid \right);$$
 \item The \emph{passive coefficients} of $x$ is the finite vector 
   $$u_p = \left( k ~ \big|~ k \in \mathds{N} \textrm{ and } 
u^k \in ]m;M[ \textrm{ and }  k \le \mid x \mid \right).$$
 \end{itemize}
 \end{definition}

For a given host content $x$,
MSCs are then ranks of $x$  that describe the relevant part
of the image, whereas LSCs translate its less significant parts.

\begin{remark}
When MSCs and LSCs represent a sequence of
bits, they are also called Most Significant Bits (MSBs) and Least Significant
Bits (LSBs). In the rest of this article, the two notations  will be used
depending on the context.
\end{remark}

\begin{example}
These two definitions are illustrated on Figure~\ref{fig:MSCLSC}, where the
significance function $(u^k)$ is defined as in Example~\ref{Exemple LSC}, $m=5$,
and $M=6$.

\begin{figure}[htb]

\begin{minipage}[b]{.98\linewidth}
  \centering
  \centerline{\includegraphics[width=2.cm]{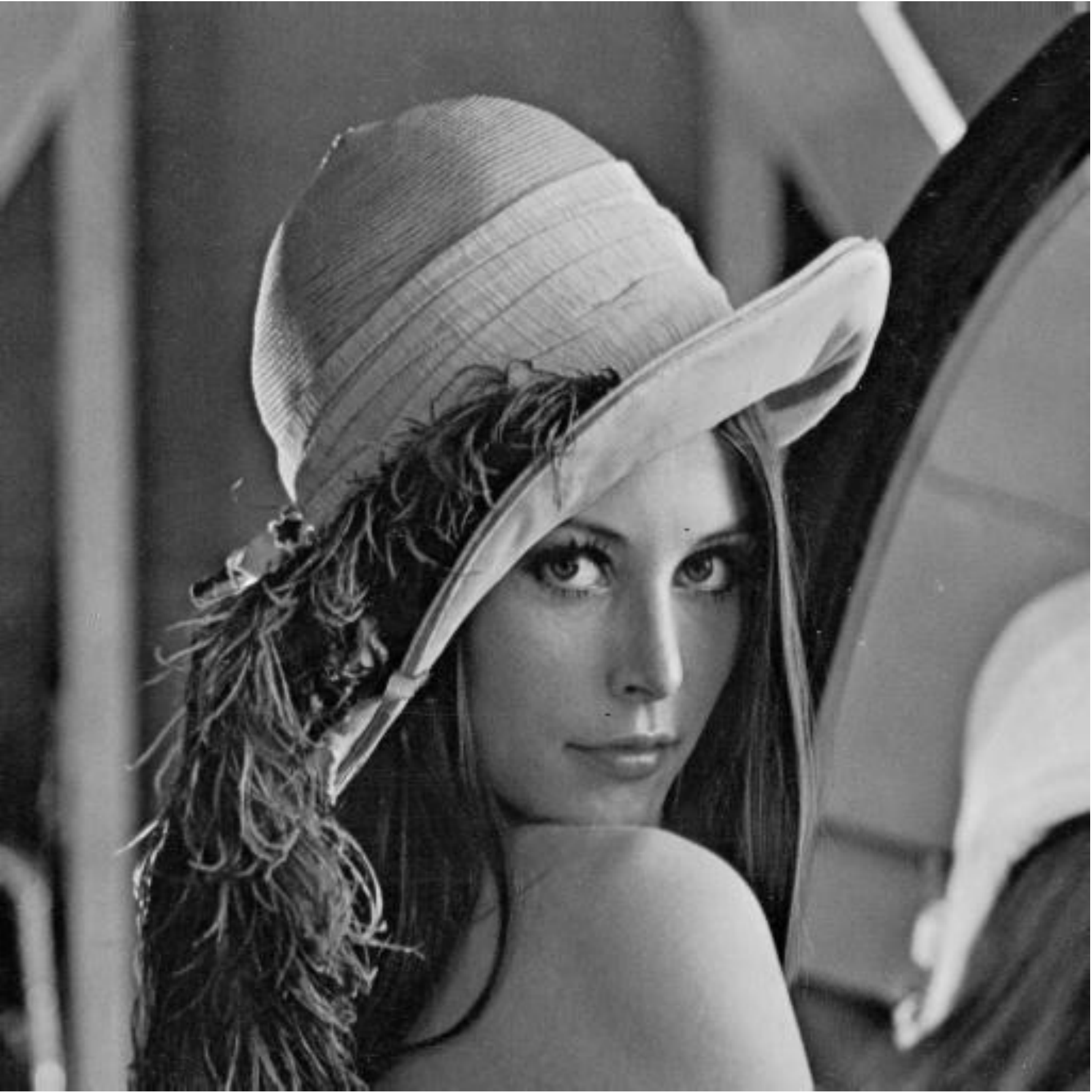}}
  \centerline{(a) Original Lena}
\end{minipage}
\begin{minipage}[b]{.49\linewidth}
  \centering
    \centerline{\includegraphics[width=2.cm]{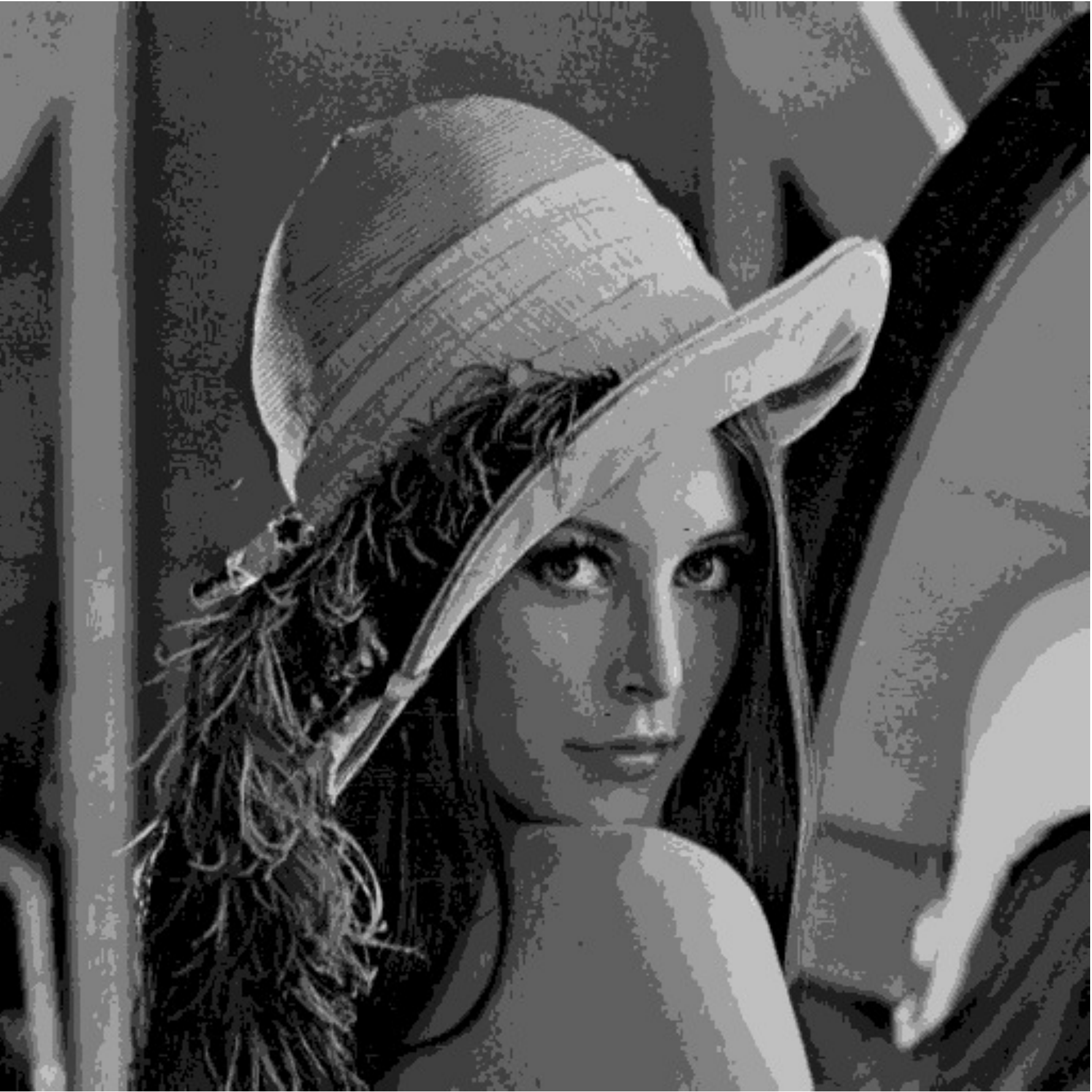}}
  \centerline{(b) MSCs of Lena}
\end{minipage}
\hfill
\begin{minipage}[b]{0.49\linewidth}
  \centering
    \centerline{\includegraphics[width=2.cm]{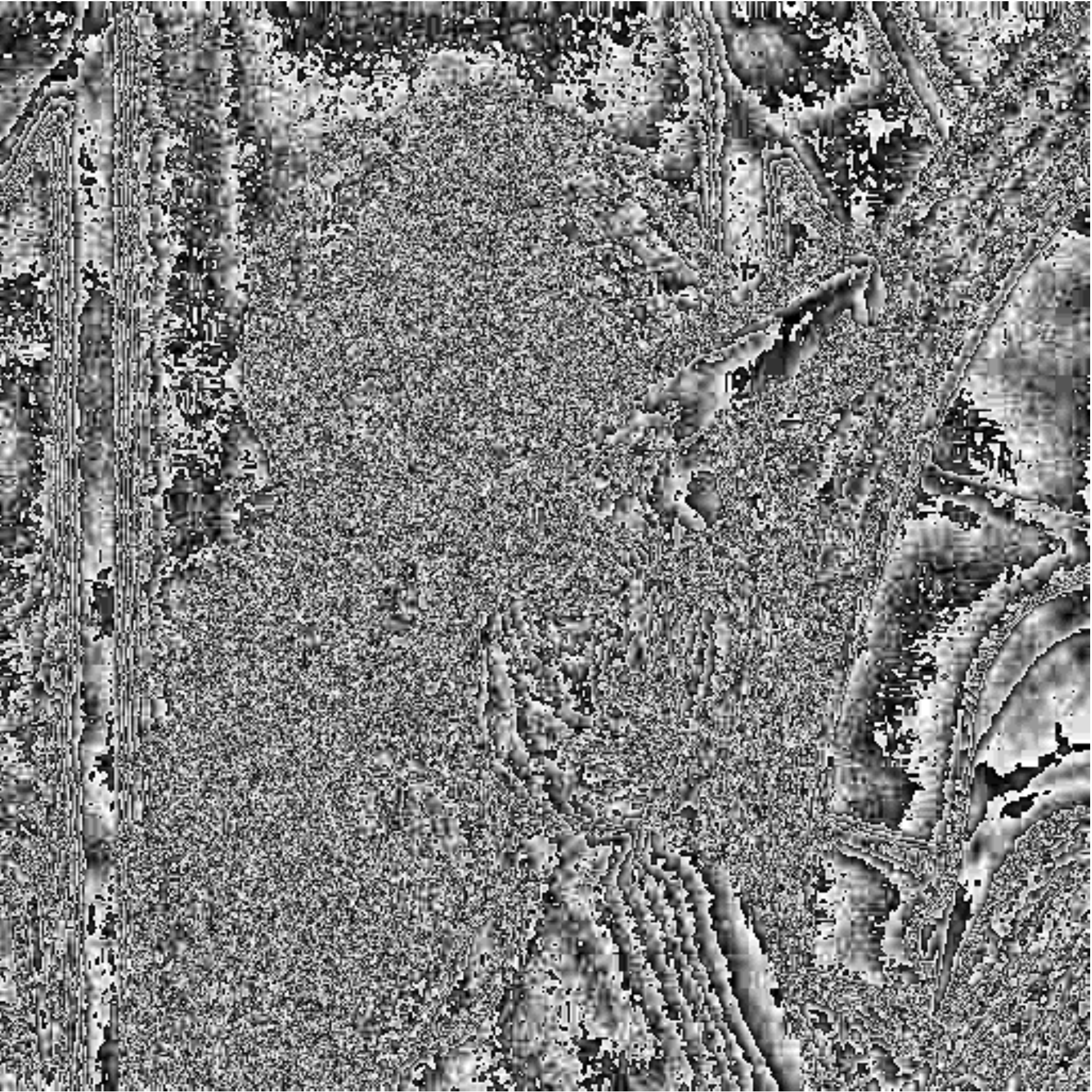}}
  
  \centerline{(c) LSCs of Lena ($\times 17$)}
\end{minipage}
\caption{Most and least significant coefficients of Lena}
\label{fig:MSCLSC}
\end{figure}
\end{example}


\section{Steganographic schemes}

\label{stateArt}
To compare the approach with other schemes, we now present recent steganographic
approaches, namely YASS (Cf setc.~\ref{sec:yass}), nsF5 (Cf
setc.~\ref{sec:nsf5}), MMx (Cf setc.~\ref{sec:mmx}), and HUGO (Cf
setc.~\ref{sec:hugo}). One should find more details in~\cite{Fridrich:2009:SDM:1721894}.
 
\subsection{YASS}\label{sec:yass}
YASS (\emph{Yet Another Steganographic Scheme})~\cite{DBLP:conf/ih/SolankiSM07} is a 
steganographic approach dedicated to JPEG cover. 
The main idea of this algorithm is to hide data 
into $8\times 8$ randomly chosen inside $B\times B$ blocks 
(where $B$ is greater than 8) instead of choosing standard 
$8\times 8$ grids used  by JPEG compression.
The self-calibration process commonly embedded into blind steganalysis schemes
is then confused by the approach.
In the paper~\cite{SSM09}, further variants of YASS have been proposed
simultaneously to enlarge the embedding rate and to improve the 
randomization step of block selecting.
More precisely, let be given a message $m$ to hide, a size $B$, $B \ge 8$, 
of blocks. 
The YASS algorithm follows. 
\begin{enumerate}
\item Computation of $m'$, which is 
  the Repeat-Accumulate error correction code of $m$.
\item In each big block of size $B \times B$ of cover, successively do:
  \begin{enumerate}
\item Random selection of an $8 \times 8$ block $b$ using w.r.t. a secret key. 
\item Two-dimensional DCT transformation of $b$ and normalisation of coefficient
  w.r.t a  predefined quantization table. 
  Matrix is further referred to as $b'$. 
\item A fragment of $m'$ is embedded into some LSB of $b'$. Let $b''$ be the
resulting matrix.
\item The matrix $b''$ is  decompressed back to the spatial domain leading to a new $B \times B$ block.
\end{enumerate}
\end{enumerate}

\subsection{nsF5}\label{sec:nsf5}
The nsF5 algorithm~\cite{DBLP:conf/mmsec/FridrichPK07} extends the F5 
algorithm~\cite{DBLP:conf/ih/Westfeld01}.
Let us first have a closer look  on this latter.

First of all, as far as we know, F5 is the first steganographic approach 
that solves the problem of remaining unchanged a part (often the end) 
of the file. 
To achieve this, a subset of all the LSB is computed thanks to a 
pseudo random number generator seeded with a user defined key. 
Next, this subset is split into blocks of $x$ bits.
The algorithm takes benefit of binary matrix embedding to
increase it efficiency. 
Let us explain this embedding on a small illustrative example where 
a part $m$ of the message  has to be embedded into this $x$ LSB of pixels   
which are respectively  a 3 bits column vector and a 7 bits column vector. 
Let then $H$ be the binary Hamming matrix  
$$
H = \left(
\begin{array}{lllllll}
 0 & 0 & 0 & 1 & 1 & 1 & 1 \\
 0 & 1 & 1 & 0 & 0 & 1 & 1 \\
 1 & 0 & 1 & 0 & 1 & 0 & 1 
\end{array}
\right)
$$
The objective is to modify $x$ to get $y$ s.t. $m = Hy$.
In this algebra, the sum and the product respectively correspond to
the exclusive \emph{or} and to the \emph{and} Boolean operators.
If $Hx$ is already equal to $m$, nothing has to be changed and $x$ can be sent.
Otherwise we consider the difference $\delta = d(m,Hx)$ which is expressed 
as a vector : 
$$
\delta = \left( \begin{array}{l}
\delta_1 \\
\delta_2 \\
\delta_3
\end{array} 
\right)  
\textrm{ where $\delta_i$ is 0 if $m_i = Hx_i$ and 1 otherwise.} 
$$
Let us thus consider the $j$th column of $H$ which is equal to $\delta$.   
We denote by $\overline{x}^j$ the vector  we obtain by
switching the $j$th component of $x$, 
that is, $\overline{x}^j = (x_1 , \ldots, \overline{x_j},\ldots, x_n )$.
It is not hard to see that if $y$ is $\overline{x}^j$, then 
$m = Hy$.
It is then possible to embed 3 bits in only 7 LSB of pixels by modifying 
on average $1-2^3$ changes. 
More generally, the F5 embedding efficiency should theoretically be 
$\frac{p}{1-2^p}$.

However, the event when the coefficient resulting from this LSB switch
becomes zero (usually referred to as \emph{shrinkage}) may occur.
In that case, the recipient cannot determine 
whether the coefficient was -1, +1 and has changed to 0 due to the algorithm or 
was  initially 0. 
The F5 scheme solves this problem first by defining a LSB  
with the  following (not even) function:
$$
LSB(x) = \left \{ 
\begin{array}{l}
1 - x \mod 2 \textrm{ if }  x< 0 \\
x \mod 2 \textrm{ otherwise.}
\end{array}
\right..
$$
Next, if the coefficient has to be changed to 0, the same bit message 
is re-embedded in the next group of $x$ coefficient LSB.

The scheme nsF5 focuses on steps of Hamming coding and ad'hoc shrinkage 
removing. It replaces them with a \emph{wet paper code} approach 
that is based on a random binary matrix. More precisely,
let $D$ be a random binary matrix of size $x \times n$ without replicate nor 
null columns: consider for instance a subset of $\{1, 2^x\}$ of cardinality $n$ 
and write them as binary numbers. The subset is generated thanks to a PRNG 
seeded with a shared key. In this block of size $x$, one choose to embed only 
$k$ elements of the message $m$. By abuse, the restriction of the message is again called $m$. It thus remains $x-k$ (wet) indexes/places where 
the information shouldn't be stored. Such indexes are generated too with the 
keyed PRNG. Let $v$ be defined by the following equation: 
 
 \begin{equation}
  Dv = \delta(m,Dx).
 \end{equation}
This equation may be solved by Gaussian reduction or other more efficient
algorithms. If there is a solution, one have the list of indexes to modify 
into the cover. The nsF5 scheme implements such a optimized algorithm  
that is to say the LT codes.

\subsection{MMx}\label{sec:mmx}
Basically, the MMx algorithm~\cite{DBLP:conf/ih/KimDR06} embeds message 
in a selected set of LSB cover coefficients using Hamming
codes as the F5 scheme. However,
instead of reducing as many as possible the number of modified elements, 
this scheme aims at reducing the embedding impact. To achieve this it allows 
to modify more than one element if this leads to decrease distortion.

Let us start again with an example with a $[7,4]$ Hamming codes, \textit{i.e},
let us embed 3 bits into 7 DCT coefficients, $D_1, \ldots, D_7$. 
Without details, let $\rho_1, \ldots, \rho_7$ be the embedding impact whilst 
modifying coefficients $D_1, \ldots, D_7$ (see~\cite{DBLP:conf/ih/KimDR06} 
for a formal definition of $\rho$). Modifying element at index $j$ 
leads to a distortion equal to $\rho_j$. However, instead of switching 
the value at index $j$, one should consider to find all other columns  
of $H$, $j_1$, $j_2$ for instances, s.t. the sum of them 
is equal to the $j$th column and to compare $\rho_j$ with 
$\rho_{j_1} + \rho_{j_2}$. If one of these sums is less than $\rho_j$,
the sender has to change these coefficients instead of the $j$ one.
The number of searched indexes (2 for the previous example) gives the name 
of the algorithm. For instance in MM3, one check whether the message can be 
embedded by modifying 3 pixel or less each time.

\subsection{HUGO}\label{sec:hugo}
The HUGO~\cite{DBLP:conf/ih/PevnyFB10} steganographic scheme is mainly designed to minimize 
distortion caused by embedding. To achieve this, it is firstly based on an
image model given as SPAM~\cite{DBLP:journals/tifs/PevnyBF10}
features and next integrates image
correction to reduce much more distortion. 
What follows refers to these two steps.

The former first computes the SPAM features. Such  calculi
synthesize the probabilities that the difference  
between consecutive horizontal (resp. vertical, diagonal) pixels 
belongs in a set of pixel values which are closed to the current pixel value
and whose radius is a parameter of the approach.  
Thus, a fisher linear discriminant method defines the radius and  
chooses between directions (horizontal, vertical, etc.) of analyzed pixels
that gives the best separator for detecting embedding changes.
With such instantiated coefficients, HUGO can synthesize the embedding cost 
as a function $D(X,Y)$ that evaluates distortions between $X$ and $Y$. 
Then  HUGO computes the matrices of 
$\rho_{i,j} = \max(D(X,X^{(i,j)+})_{i,j}, D(X,X^{(i,j)-})_{i,j})$ 
such that $X^{(i,j)+}$ (resp. $X^{(i,j)-}$ ) is the cover image $X$ where 
the the $(i,j)$th pixel has been increased (resp. has been decreased) of 1.

The order of modifying pixel is critical: HUGO surprisingly 
modifies pixels in decreasing order of $\rho_{i,j}$. 
Starting with $Y=X$, it increases or decreases its $(i,j)$th pixel
to get the minimal value of 
$D(Y,Y^{(i,j)+})_{i,j}$ and  $D(Y,Y^{(i,j)-})_{i,j}$. 
The matrix $Y$ is thus updated  at each round.

\section{The new steganographic
process $\mathcal{DI}_3$}
\subsection{Implementation}\label{sec:algo-di3}

%
In this section, a new algorithm which is inspired from the
schemes $\mathcal{CIW}_1$ and $\mathcal{CIS}_2$ respectively described
in~\cite{fgb11:ip} and~\cite{gfb10:ip} is presented. Compare to the first one,
it is a steganographic scheme, not just a watermarking technique.  Unlike
$\mathcal{CIS}_2$ which require embedding keys with three strategies, only one
is required for $\mathcal{DI}_3$. So compare
to $\mathcal{CIS}_2$ which is also a steganographic process, it is easier to
implement for Internet applications especially in order to contribute to a semantic web. Moreover,
since $\mathcal{DI}_3$ is a particular instance of $\mathcal{CIS}_2$, it is
clearly faster than this one because
in $\mathcal{DI}_3$ there is no operation to mix the message on the contrary
on the initial scheme. The fast execution of such an algorithm is critical for
internet applications.

In the following algorithms, the following notations are used:
\begin{notation}
$S$ denotes the embedding and extraction strategy, 
 $H$ the host content or the stego-content depending of the context.
  $LSC$ denotes the old or new
LSCs of the host or stego-content $H$ depending of the context too.
 $N$ denotes the number of LSCs,
 $\lambda$ the number of iterations to realize,
 $M$ the secret message, and
 $P$ the width of the message (number of bits).
\end{notation}

Our new scheme theoretically presented in~\cite{ihtiap-2012} is here described
by three main algorithms:
\begin{enumerate}
  \item The first one, detailed in Algorithm~\ref{algo:strategy}
allows to generate the embedding strategy of the system which
is a part of the embedding key in addition with the choice of the LSCs and
the number of iterations to realize.

\item The second one, detailed
in Algorithm~\ref{algo:embed} allows to embed the message into
the LSCs of the cover media using the strategy. The 
strategy has been generated by the first algorithm and the same number of
iterations is used.
\item The last one, detailed in Algorithm~~\ref{algo:extract} allows to extract
the secret message from the LSCs of the media (the stego-content) using the
strategy wich is a part of the extraction key in addition with the width of the
message.
\end{enumerate}   

In adjunction of these three functions, two other complementary functions have
to be used:

\begin{enumerate}
  \item The first one, detailed in Algorithm~\ref{algo:signification-function},
  allow to extract MSCs, LSCs, and passive coefficients from the host content.
  Its implementation is based on the concept of signification
  function described in Definition~\ref{def:msc-lsc}.
 \item The last one, detailed in Algorithm~\ref{algo:build-function}, allow to
 rebuild the new host content (the stego-content) from the corresponding MSCs,
 LSCs, and passive coefficients. Its implementation is also based on the concept
 of signification function described in Definition~\ref{def:msc-lsc}. This
 function realize the invert operation of the previous one.
\end{enumerate}

\begin{remark}
The two previous algorithms have
to be implemented by the user depending on each application context should be
adjusted accordingly: either in spatial description, in frequency description, or in other description. They
correspond to the theoretical concept described in Definition~\ref{def:msc-lsc}. Their
implementation depends on the application context.
\end{remark}
\begin{example}
For example the algorithm~\ref{algo:signification-function} in spatial domain
can correspond to the extraction of the 3 last bits of each pixel as LSCs, the 3
first bits as MSCs, and the 2 center bits as passive coefficients.
\end{example}

%
%

\begin{algorithm}[h]
\tcc{$S$ is a sequence of
integers into $\llbracket 0,P-1 \rrbracket$, such that $(S_{n_0},\ldots,S_{n_0+P-1})$ is injective on
$\llbracket 0,P-1 \rrbracket$.}
\KwResult{$S$: The strategy, integer sequence $(S_0,S_1,\ldots)$.}
\Begin{
$n_0 \longleftarrow L - P + 1$\;
\If{$P > N$ OR $n_0 < 0$}{
\Return{ERROR}}
$S \longleftarrow$ Array of width $\lambda$, all values initialized to 0\;
$cpt \longleftarrow 0$\;
\While{$cpt < n_0$}{
$S_{cpt} \longleftarrow $Random integer in $\llbracket 0,P-1
\rrbracket$.\;
$cpt \longleftarrow cpt + 1$\;}
$A \longleftarrow$ We generate an arrangement of $\llbracket 0,P-1
\rrbracket$\;
\For{$k \in \llbracket 0,P-1\rrbracket$}{
$S_{n_0 + k} \longleftarrow A_k$\;
}
\Return{$S$}
}
\caption{$strategy(N,P,\lambda)$}
\label{algo:strategy}
\end{algorithm}

\begin{algorithm}[h]
\KwResult{New LSCs with embedded message.}
\Begin{
$N \longleftarrow$ Number of LSCs in $LSC$\;
$P \longleftarrow$ Width of the message $M$\;
\For{$k \in \llbracket 0,\lambda\rrbracket$}{
$i \longleftarrow S_k$\;
$LSC_{i} \longleftarrow M_i$\;
}
\Return{$LSC$}
}
\caption{$embed(LSC, M, S, \lambda)$}
\label{algo:embed}
\end{algorithm}

\begin{algorithm}[h]
\KwResult{The message to extract from $LSC$.}
\Begin{
$RS \longleftarrow$ The strategy $S$ written in reverse order.\;
$M \longleftarrow$ Array of width $P$, all values initialized to 0\;
\For{$k \in \llbracket 0,\lambda\rrbracket$}{
$i \longleftarrow RS_k$\;
$M_{i} \longleftarrow LSC_i$\;
}
\Return{$M$}
}
\caption{$extract(LSC, S, \lambda,P)$}
\label{algo:extract}
\end{algorithm}

\begin{algorithm}[h]
\KwData{$H$: The original host content.}
\KwResult{$MSC$: MSCs of the host content $H$.}
\KwResult{$PC$: Passive coefficients of the host content $H$.}
\KwResult{$LSC$: LSCs of the host content $H$.}
\Begin{
\tcc{Implemented by the user.}
\Return{$(MSC,PC,LSC)$}
}
\caption{$significationFunction(H)$}
\label{algo:signification-function}
\end{algorithm}

\begin{algorithm}[h]
\KwResult{$H$: The new rebuilt host content.}
\Begin{
\tcc{Implemented by the user.}
\Return{$(MSC,PC,LSC)$}
}
\caption{$buildFunction(MSC,PC,LSC)$ )\label{algo:build-function}}
\end{algorithm}



\subsection{Discussion}

We first notice that our $\mathcal{DI}_3$ scheme embeds the message in LSB as
all the other approaches.
Furthermore, among all the LSB, the choice of those which are modified
according to the message is based on a secured PRNG whereas F5, and thus nsF5
only require a PRNG.
Finally in this scheme, we have postponed the optimization of considering
again a subset of them according to the distortion their modification
may induce. According to us, further theoretical study
are necessary to take this feature into consideration.
In future work, it is planed to compare the robustness and efficiency of all the
schemes in the context of semantic web. To initiate this study in this first
article, the robustness of $\mathcal{DI}_3$ is detailled in the next section.

\section{Robustness Study}\label{sec:robustness-study}

This section evaluates the robustness of our approach~\cite{bcg11:ij}.

Each experiment is build on a set of 50 images which are randomly selected
among database taken from the BOSS contest~\cite{DBLP:conf/ih/BasFP11}. 
Each cover is a $512\times 512$ greyscale digital image. 
The relative payload
is always set with 0.1 bit per pixel. Under that constrain,
the embedded message $m$ is a sequence of 26214 randomly generated bits. 

Following the same model of robustness studies in previous similar work in the
field of information hiding, we choose some classical attacks like cropping,
compression, and rotation studied in this research work. Other attacks
and geometric transformations will be explore in a complementary study. Testing
the robustness of the approach is achieved by successively applying on stego content images attacks. Differences between the message that is extracted from the attacked image and the original one are computed and expressed as percentage.

\begin{figure*}[htb]

\begin{minipage}[b]{.24\linewidth}
  \centering
    \centerline{\includegraphics[width=5cm]{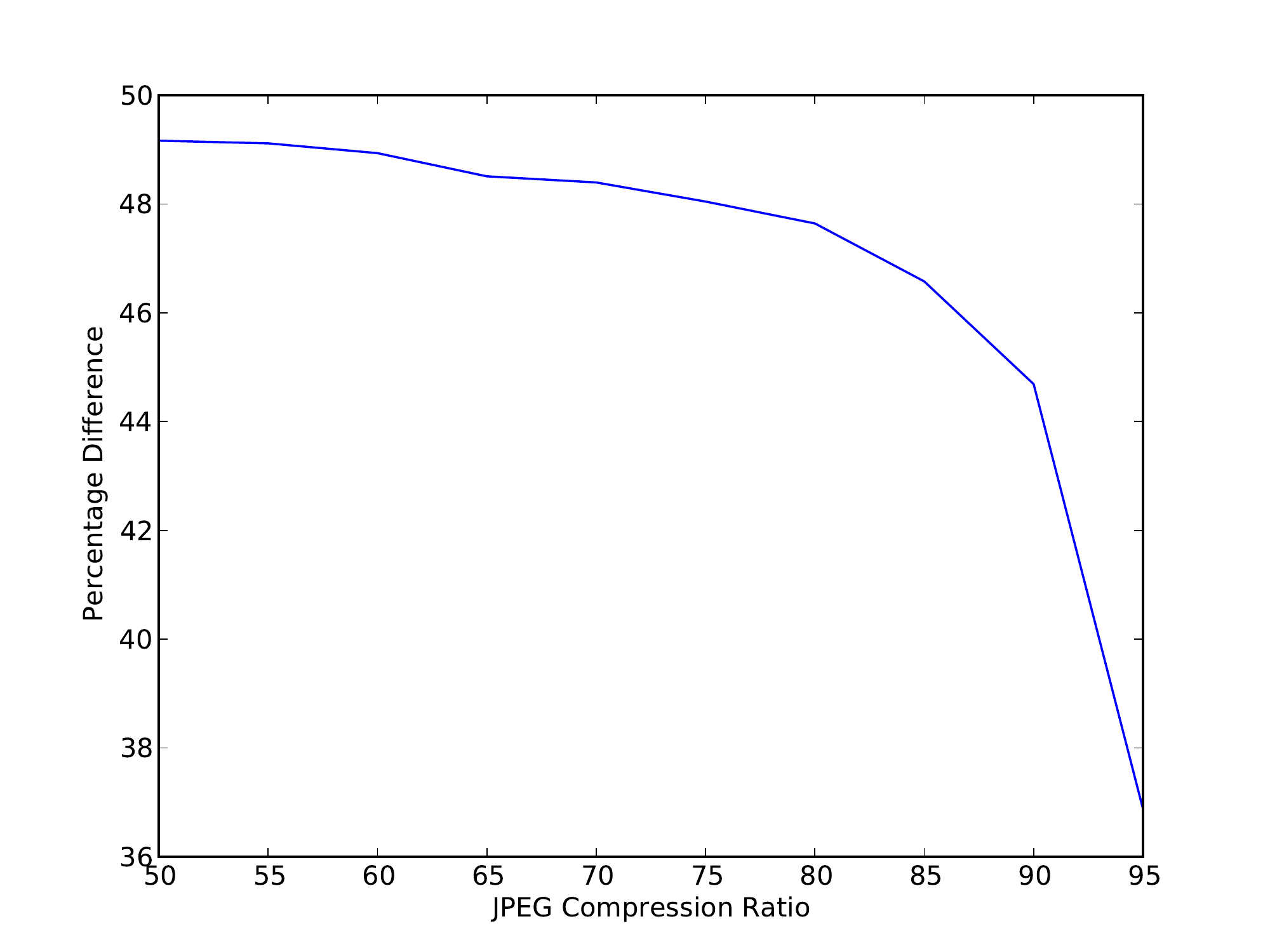}}
    \centerline{(a) JPEG effect.}
\end{minipage}
\hfill
\begin{minipage}[b]{0.24\linewidth}
  \centering
    \centerline{\includegraphics[width=5cm]{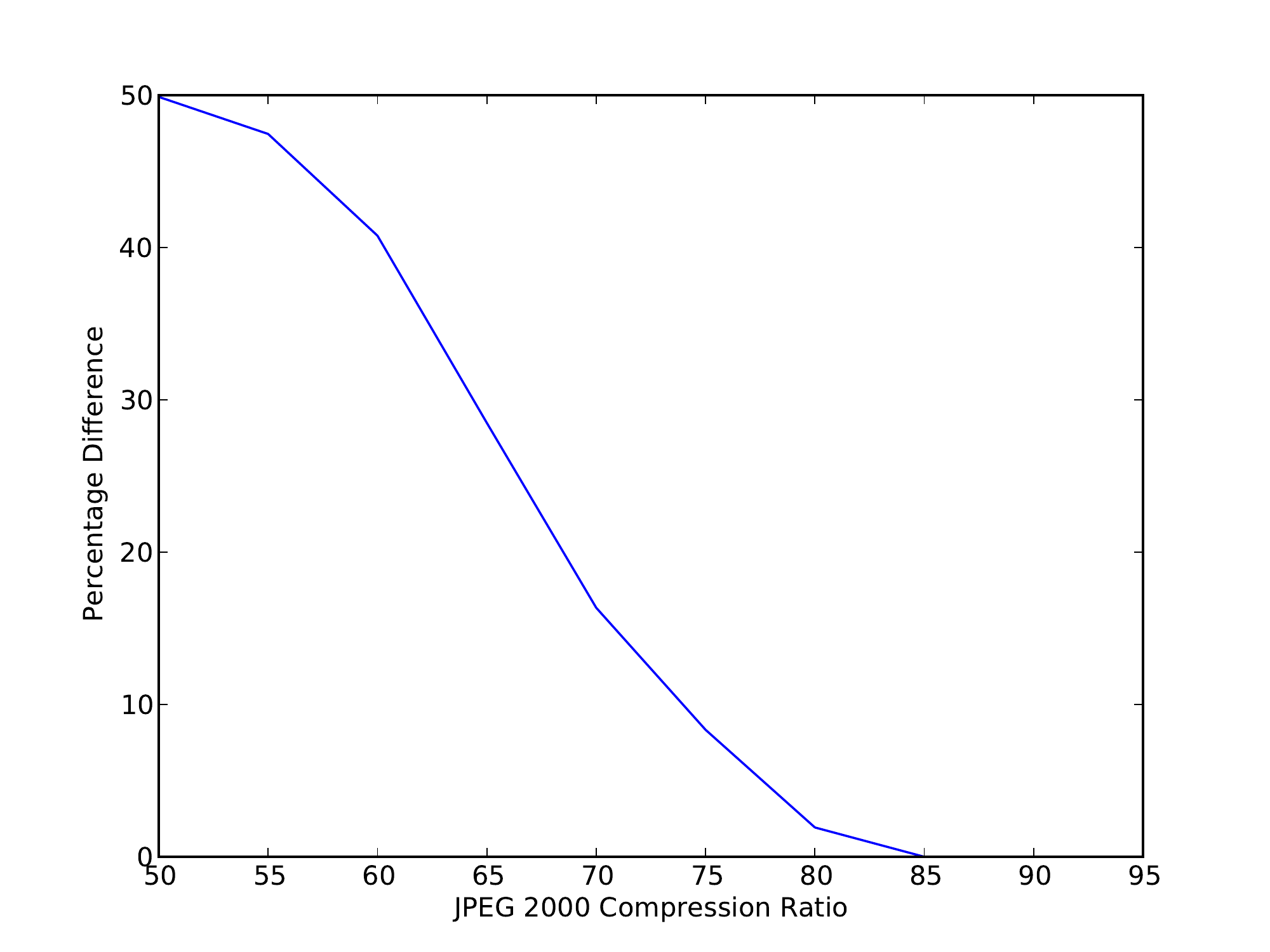}}
    \centerline{(b) JPEG 2000 effect.}
\end{minipage}
\hfill
\begin{minipage}[b]{.24\linewidth}
  \centering
    \centerline{\includegraphics[width=5cm]{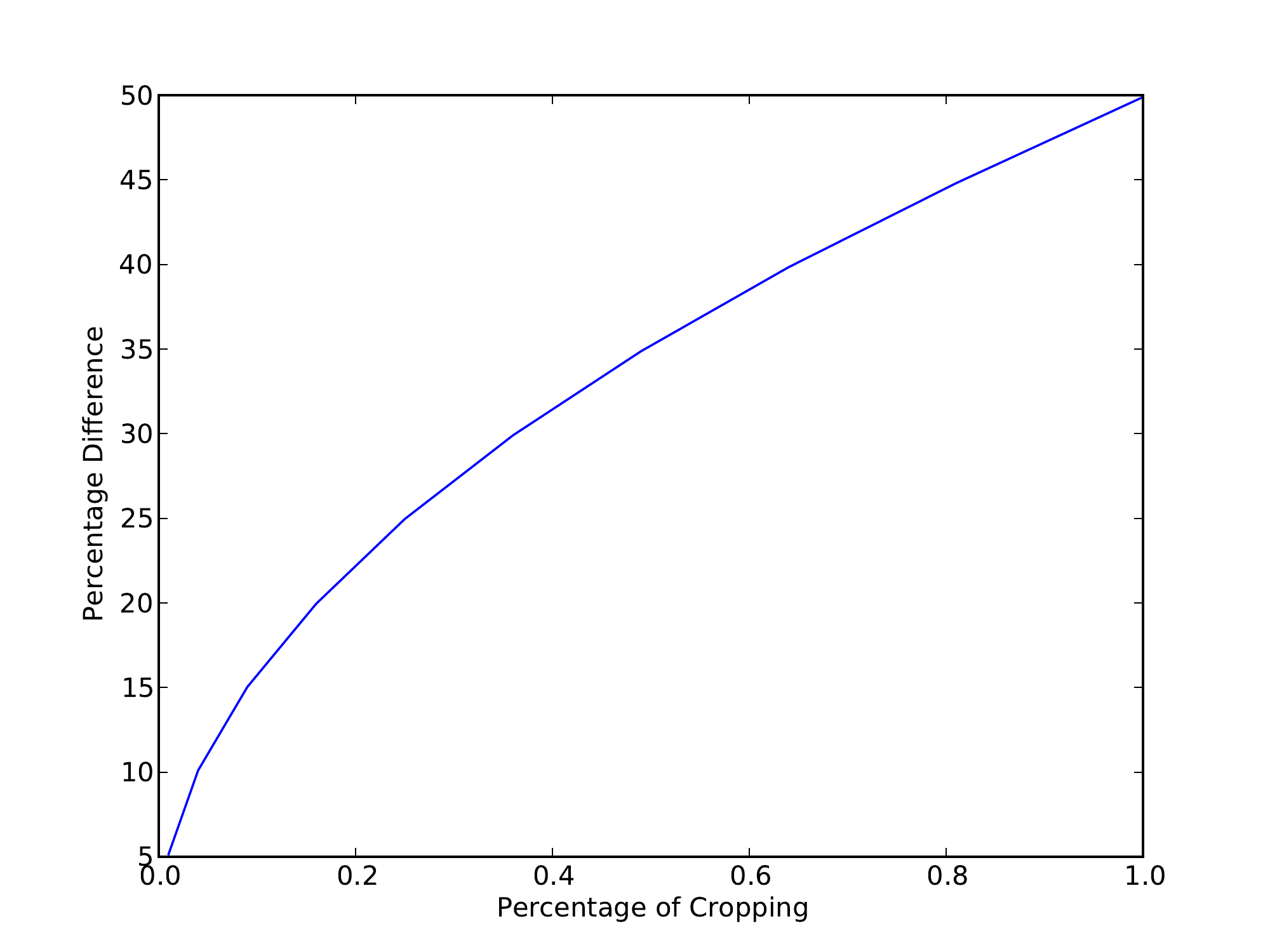}}
    \centerline{(c) Cropping attack.}
\end{minipage}
\hfill
\begin{minipage}[b]{0.24\linewidth}
  \centering
    \centerline{\includegraphics[width=5cm]{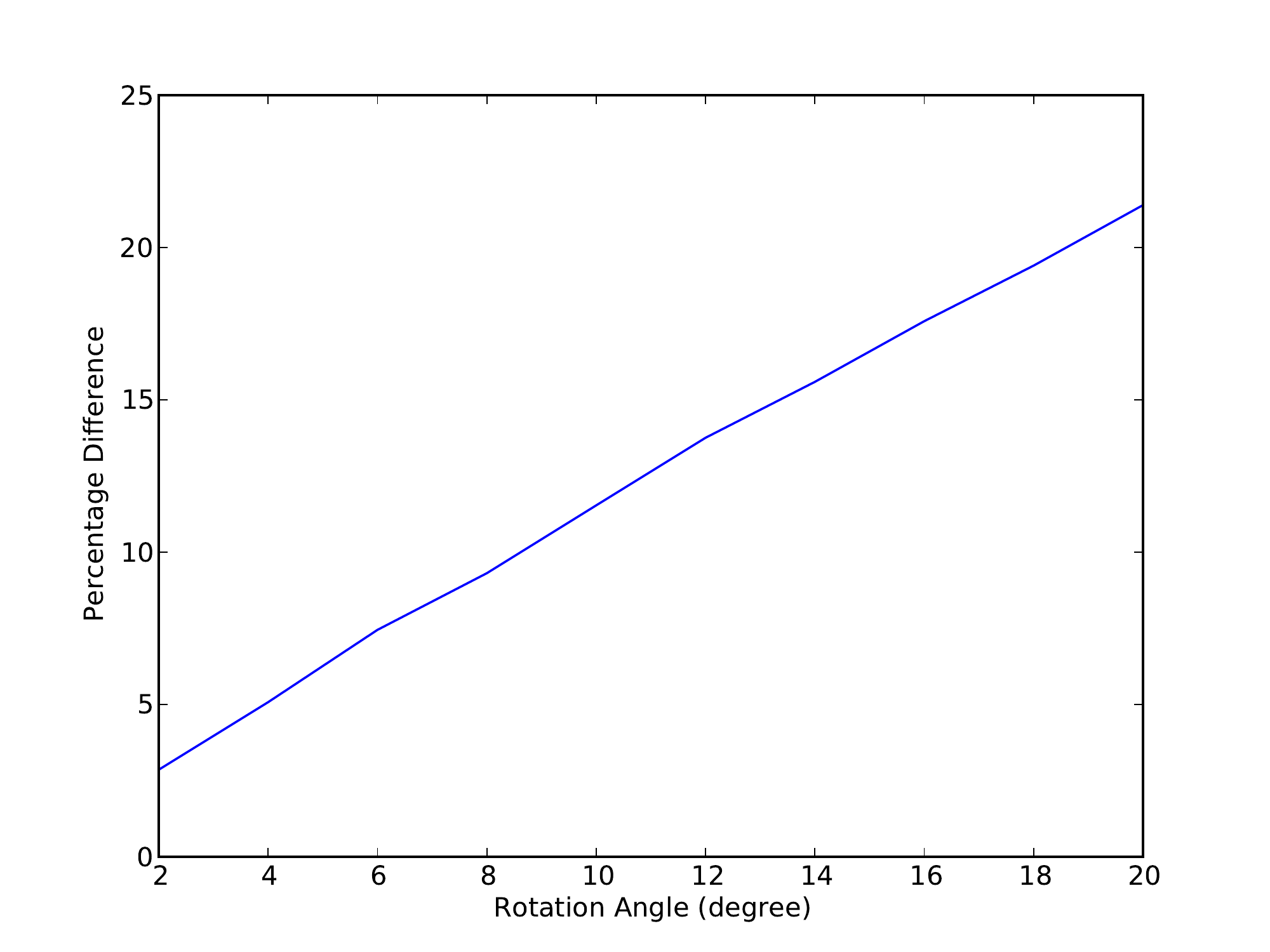}}
    \centerline{(d) Rotation attack.}
\end{minipage}
\caption{Robustness of $\mathcal{DI}_3$ scheme facing several attacks (50 images
from the BOSS repository)}
\label{fig:robustness-results}
\end{figure*}

To deal with cropping attack, different percentage of cropping
(from 1\% to 81\%) are applied on the stego content image.  
Fig.~\ref{fig:robustness-results}~(c) presents effects of such an attack.

%

We address robustness against JPEG an JPEG 2000 compression.
Results are respectively presented in Fig.~\ref{fig:robustness-results}~(a) and
in Fig.~\ref{fig:robustness-results}~(b).

Attacked based on geometric transformations are addressed through  
rotation attacks: two opposite rotations 
of angle $\theta$ are successively applied around the center of the image.
In these geometric transformations, angles range from 2 to 20 
degrees.  
Results  effects of such an attack are also presented in
Fig.~\ref{fig:robustness-results}~(d).


From all these experiments, one firstly can conclude that
the steganographic scheme does not present obvious drawback and
resists to all the attacks:
all the percentage differences are so far less than 50\%.

The comparison with robustness of other steganographic schemes exposed in the
work will be realize in a complementary study, and the best utilization of each
one in several context will be discuss.

\section{Conclusion and future work}\label{sec:conclusion-future-works}

In this research work, a new information hiding algorithm has been introduced to
contribute to the semantic web. We have focused our work on the robustness
aspect. The  security has been studied in an other
work~\cite{ihtiap-2012}. Even if this new scheme $\mathcal{DI}_3$
does not possess topological properties (unlike the
$\mathcal{CIS}_2$~\cite{fgb11:ip}), its level of security seems to be sufficient for
Internet applications. Particularly in the framework of the semantic web it is
required to have robust steganographic processes. The security aspects is less
important in this context. Indeed, it is important that the enrichment
information persist after an attack.
Especially for JPEG 2000 attacks,
which are the two major attacks used in an internet framework. Additionally,
this new scheme is faster than $\mathcal{CIS}_2$. This is a major advantage for
an utilization through the Internet, to respect response times of web sites.



In a future work we intend to prove rigorously that $\mathcal{DI}_3$ is not
topologically secure. The tests of robustness will be realized on a larger set of images of different
  types and sizes, using resources of the \emph{Mésocentre de calcul de
  Franche-Comt\'{e}~\cite{www:mesocenter-ufc} (an High-Performance Computing
  (HPC) center)} and using Jace environment~\cite{bhm09:ip}, to take benefits
  of parallelism. So, the robustness and efficiency of our scheme
  $\mathcal{DI}_3$ will be compared to other schemes in order to show the best utilization in several
  contexts. Other kinds of attacks will be explored to evaluate more
  completely the robustness of the proposed scheme. For instance, robustness of the $\mathcal{DI}_3$ against Gaussian blur, rotation, contrast, and zeroing attacks will be regarded, and compared with a
larger set of existing steganographic schemes as those described in this
article. Unfortunately these academic algorithms are mainly designed to show their ability in embedding. Decoding aspect is rarely treated, and rarely implemented
at all. Finally, a first web search engine compatible with the proposed robust watermarking scheme
will be written, and automatic tagging of materials found on the Internet will be realized,
to show the effectiveness of the approach.

%
%
%
%
%

\bibliographystyle{plain}
\bibliography{jabref}

\end{document}